Invited Review

# Raman Optothermal Technique for Measuring Thermal Conductivity of Graphene and Related Materials

Hoda Malekpour and Alexander A. Balandin*

Department of Electrical and Computer Engineering, University of California, Riverside, California, USA.

*Abstract*

**We describe Raman spectroscopy based method of measuring thermal conductivity of thin films, and review significant results achieved with this technique pertinent to graphene and other two-dimensional materials. The optothermal Raman method was instrumental for the discovery of unique heat conduction properties of graphene. In this method, Raman spectroscopy is used to determine the local temperature of the sample while the excitation laser is utilized as a heat source. The success of Raman spectroscopy in investigating thermal conductivity of suspended graphene and graphene-based thin films motivated extension of this technique to other materials systems and films.**

## 1. Introduction: Raman Spectroscopy of Graphene

Raman spectroscopy is a powerful tool for characterizing graphene-based films. The Raman spectrum of single layer graphene (SLG) consists of several well resolved bands.[1] The G-band corresponds to the $E_{2G}$ vibrational mode of the Brillouin zone (BZ) center phonons. The D-band, the defect or disorder related mode, is attributed to the breathing mode of a carbon ring and originates from the resonant inter-valley process at the BZ edge.[2-5] This band appears in graphene with defects owing to relaxation of the phonon momentum conservation rules. The D′ band corresponds to the intra-valley resonant process. The 2D band, an overtone of the D band, is always present in Raman spectrum of graphene even without defects. Fig 1(a) shows the Raman spectrum and the distinct peaks for pristine and defective graphene.[6] While the G band position does not depend on the excitation energy ($E_L$), the D and 2D band are strongly dispersive with $E_L$. The latter originates from the Kohn anomaly at $K$ point of BZ.[7]

Raman spectrum of graphene and few layer graphene (FLG) carries important characterization information of the materials. As the number of graphene layers ($N$) increases the Raman spectrum evolves, creating two extra modes: the sheer mode[8] and layer breathing mode (LBMs)[9-11] both originating from relative motion of graphene planes.[1] While the appearance of the sheer mode is a direct sign of the presence of more than one layer, its low wavenumber[8,12] ($< 50\ cm^{-1}$) stands below the cutoff frequency of most notch and edge filters.[1] However, owing to the strong correlation with the in-plane vibrational modes (G and 2D bands), the number of layers can be accurately detected using these modes.[13-18]

Both the shape and intensity of 2D band evolve with increasing $N$ owing to the modification of electronic bands.[14] In addition to the number of layers, this evolution could be used to detect the relative orientation of the layers.[19] The G band intensity on the other hand scales up with increasing $N$ enabling one to probe the layer numbers by monitoring the Raman G to 2D band intensity ratio.[13] Fig 1(b) shows the evolution of graphene's visible Raman spectrum with $N$.[13] While visible Raman is the most common method for probing number of layers in FLG,[14-18] the UV excitation Raman can provide additional information for verifying $N$.[13]

Raman spectroscopy is also known as a powerful tool for quantifying the amount of defects.[20] A three-stage classification of defect regimes was introduced by Ferrari et al[3] explaining the changes in the Raman spectrum of graphite moving toward amorphous carbon. A set of empirical relations was introduced to directly quantify the amount of defects in graphene for the low defect density regimes.[20] The Raman D to G band intensity ratio ($I_D/I_G$) is reported to not only depends on the excitation energy,[3,4,20,21,22] but also sensitive to the inter-defect distance ($L_D$).[3,20] For the low defect density regime, this ratio is proportional to ${L_D}^2$.[3] One should note that not all types of defects could be detected by this technique, and only the defects that contribute to the inter-valley D process would give rise to the D-band intensity. Raman spectroscopy can also be sensitive to the nature of defects. It was reported that Raman D to D′ band intensity ratio could be used to characterize the nature of defects.[23] The maximum value of this intensity ratio was reported for $sp^3$-defects ($I_D/I_{D'}$ $\sim$13). For vacancy type defects this intensity is reduced to $\sim$7 and reaches its minimum $\sim$3.5 for boundary type defects.[23] In addition to the number[13-18] and orientation of graphene layers,[19] as well as the quality and quantity of defects,[3,20,24] Raman spectroscopy is an excellent tool for investigating strain, [25-31] doping,[32-40] edge[41-46] and functional groups.[47-49] On the other hands, due to the temperature dependence of the Raman bands as well as the clear and distinct Raman signature of graphene, this spectroscopic technique enables one to study thermal properties of graphene-based films.[50] The subject is discussed more in detail in the following section.

**[Fig 1]**

Raman study of graphene under 488 nm excitation wavelength: (a) distinctive Raman bands of pristine and defected CVD-grown graphene. While the G and 2D bands are always present in graphene's Raman spectrum, the D and D′ bands require defects for activation. (b) The evolution of Raman spectrum obtained from the mechanically exfoliated few layer graphene with varying number of atomic layers. The intensity ratio of Raman G to 2D band is used for determining the film thickness. Fig (b) is adopted from Ref. 13 published by American Institute of Physics.

# 2. Optothermal Raman Technique for Measuring Thermal Conductivity of Graphene

Thermal conductivity is the property of a material showing how well it conducts heat. It is defined via Fourier's law, $q = -K\nabla T$, where $q$ denotes heat flux density, $K$ is thermal conductivity and $\nabla T$ shows temperature gradient. Over a large temperature range, the thermal conductivity varies with temperature, T. In anisotropic materials it is also a function of crystal orientation. Heat can be carried both by electrons and acoustic phonons and the total thermal conductivity is the sum of these two components, $K = K_p + K_e$, where $K_p$ and $K_e$ represents phonon and electron contributions, respectively. The contribution of these two parts varies in different materials. In metals, the presence of large concentration of free electrons makes the electronic part of thermal conductivity dominant. In carbon materials, the heat conduction is usually dominated by phonons. Using Wiedemann-Franz law one can extract $K_e$ having electrical conductivity of the material. The methods of measuring thermal conductivity are categorized into two groups: steady state and transient.[51] The transient group includes techniques in which the measurement of K takes place as a function of time. The "laser flash" technique is an example of this category in which thermal diffusivity ($D_T$) is measured over a large temperature range. The thermal conductivity is calculated knowing the mass density ($\rho_m$) and the specific heat ($C_p$) of the material measured independently, $K = D_T C_p \rho_m$. Another example of the transient group is the $3\omega$ method in which the thermal conductivity of a thin film is measured based on temperature dependence of electrical conductivity.[52]

In the steady state techniques, the measurement of thermal conductivity is performed independent of time. An example of this category, which is the main focus of this review paper, is optothermal Raman (OTR) technique. This technique was introduced by Balandin and co-workers for measuring the thermal conductivity of single layer graphene (SLG).[53,54] In this technique, the Raman laser light acts as a heat source to cause a local temperature rise in the suspended graphene. The local temperature rise is then measured by means of Raman thermometry. The strong temperature dependence of the Raman G peak position of graphene as well as its clear Lorentzian shape enables an accurate temperature reading. The measurement of graphene's thermal conductivity using OTR technique is performed via a two-step procedure. In the first step, the Raman G peak shift is recorded in response to increasing laser power ($\Delta P$). For this measurement the sample is suspended over a trench and connected to two heat sinks. Fig 2(a) shows the experimental sample set up initially used for measuring thermal conductivity of SLG.[55] In order to correlate the shift in G-peak ($\Delta\omega_G$) with the local temperature rise ($\Delta T$) caused by the laser heating, a calibration measurement needs to be done. In the second step, the calibration measurement, the position of the Raman G peak is recorded as a function of graphene's temperature, controlled externally. The calibration measurement in performed inside a cold/hot cell, using small excitation laser powers to avoid any laser induced heating.[53,54] The cell enables an accurate control of graphene's temperature in a wide range. Fig 2(b) shows the temperature dependence of graphene's G peak. The slope of this plot, $\chi_G$, allows one to use Raman as thermometer ($\Delta T = \Delta\omega_G/\chi_G$).

To extract the thermal conductivity values, a heat diffusion equation needs to be solved for the specific sample geometry. One of the important parameters required for this purpose is the amount of laser power absorbed by suspended graphene's film. The light absorption coefficient of graphene can be measured directly by placing a power sensor underneath the film.[6,56] It was also reported that integrated Raman intensity of G peak could be used to detect the amount of absorbed power.[53] It is important to note that the absorption coefficient of graphene strongly depends on the wavelength of incident light (Fig 2(c)).[57] Other parameters such as stress, defects, surface contamination and multiple reflections caused by substrate can influence the absorption coefficient.[6,16,58] Having the local temperature rise ($\Delta T$) in response to the increasing laser power ($\Delta P$), thermal conductivity is extracted by solving the heat diffusion equation. The laser-induced local heating inside the graphene diffuses toward the heat sinks, where the temperature of the film is held at room temperature. The heat diffusion equation can be numerically solved[6] or approximated for symmetric samples.[53]

Since the initial development of this technique by Balandin and co-workers,[53,54] the non-contact OTR has been used not only for measuring thermal conductivity of graphene based films[6,53,54,56,59,60,61] but also for a wide range of other 2D materials.[62-66] Moreover, the application of this technique is not limited to 2D structures and could be extended to relatively thick films, having enough laser power to induce local heating.[67,68] Fig 2(d) demonstrates an experimental set-up designed for measuring thermal conductivity of relatively thick films.[67] In the rest of this paper the use of OTR technique for thermal characterization of graphene and other materials in both 2D and bulk, i.e. 3D, structures will be reviewed. First, the application of OTR technique for investigating the variation of thermal conductivity with defects in 2D graphene will be discussed.

**[Fig 2]**

OTR measurement of thermal conductivity: (a) scanning electron microscopy image of micro-scale experimental setup showing suspended few layer graphene over $3$ $\mu m$-wide Si trench. (b) Temperature dependence of graphene's Raman G-band with the corresponding linear coefficient. The inset demonstrates the Raman G-band with a perfect Lorentzian shape. (c) Experimental optical absorption of graphene showing strong dependence on the excitation energy. The maximum absorption occurs at $4.62\ eV$. In addition to excitation energy, other parameters such as structural defects, bending, stress and surface contaminations might influence the absorption coefficient. (d) Optical image of the sample holder designed for performing OTR measurement on macro-scale thin films. The sample holder contains two aluminum pads for suspending the sample and serving as heat sink. The suspended graphene laminate on PET substrate is marked with an arrow. Fig (a) is adopted from Ref. 55 published by Nature Publishing Group, Fig (b) is adopted from Ref. 50 published by American Chemical Society, Fig (c) is adopted from Ref. 57 published by American Physical Society.

# 3. Optothermal Raman Investigation of Thermal Conductivity of Graphene with Defects

In this section, we review the work conducted by Malekpour et al.[6] studying the effect of defects on thermal conductivity of graphene. Graphene is well known for its superior thermal conduction properties.[53-55,69] However, its exceptionally high thermal conductivity can be deteriorated by different types of inevitable defects. These defects could be induced by polymer residue from nanofabrication,[70] the roughness of edge,[71] polycrystalline grain boundaries[72] or caused by contact with substrate or capping layer.[73-75] An example of this degradation is seen in chemical vapor deposited (CVD) graphene, always holding lower values of thermal conductivity compared to mechanically exfoliated graphene from highly ordered pyrolytic graphite (HOPG).[53,54,56,76] An additional suppression in thermal conductivity of CVD graphene can be caused possibly by the loss of polycrystalline grain orientation.[77] A lack of a quantitative experimental study on the dependence of thermal conductivity on these defects has motivated the authors to perform a thorough experimental investigation, analyzing the effect of electron beam induced defects on thermal properties of CVD graphene. The authors interpreted their experimental study using the Boltzmann transport equation and molecular dynamics simulations.[6]

The relaxation time of phonon scattering ($\tau_p$) on defects and grain boundaries is a function of phonon frequency (*f*). A change in dimensionality directly influences the phonon density of states (PDOS) leading to a different frequency dependence, $\tau_p(f)$. For example transferring from bulk 3D crystal to 2D lattice of graphene changes this dependence from $1/\tau_p \sim 1/f^4$ to $\sim 1/f^3$, [78] affecting phonon mean free path (MFP) and thus thermal conductivity. Understanding the behavior of thermal conductivity with density of defects ($N_D$) can shed light on the phonon-point defect scattering strength in two-dimensional materials. Moreover, considering the wide application of graphene and few layer graphene (FLG) in thermal management, e.g. heat spreaders[79-81] and thermal interface materials (TIMs),[82-84] the knowledge of $K(N_D)$ is critical for the development of graphene based thermal materials. This is especially important since the FLG used in thermal management is usually produced via CVD and liquid phase exfoliation (LPE) techniques, both providing graphene with large density of defects.

## a) Thermal Conductivity of CVD Graphene

For this study, CVD graphene films were grown on copper foil[85] and transferred onto a gold transmission electron microscopy (TEM) grid for the following OTR measurements. Fig. 3(a) shows the scanning electron microscopy (SEM) image of the transferred graphene on an array of square holes prepared by the TEM grid. Only the holes fully covered with graphene were selected for the thermal studies in order to avoid any complexity in $K$ extraction. The study was performed on three different covered holes to confirm the consistency and accuracy of the results. Playing the role of heat sink in OTR experimental setup, the gold TEM grid was chosen due to its high thermal conductivity ($K = 350\ W/mK$) as well as its strong attachment to graphene. Raman spectroscopy was conducted prior to the OTR measurements to ensure the crystallinity and proper quality of the transferred graphene (Fig 3(b)). The appearance of a weak D-peak in the Raman spectrum validates our earlier discussion of the presence of inevitable defects in CVD graphene.

OTR measurements were then carried out on the selected three squares of suspended graphene (SLG #1-3). The details of the OTR technique is discussed in section 2 and provided in Ref 6. For extracting the thermal conductivity, the optical absorption coefficient of graphene was measured directly by placing a photodiode power sensor underneath the sample. The absorption coefficient ($\alpha$) was measured to be ~5.7% for the laser wavelength used in the OTR experiment ($488\ nm$). This value is obviously larger than the expected absorption coefficient for the experimental excitation wavelength (Fig 2(c)) owing to possible graphene bending, surface contamination and defects induced during synthesis and transfer process.[55,86-88] Fig 3(c) shows the power dependent Raman measurement results illustrating an excellent linear fit. The temperature coefficient of G-peak was extracted to be $\chi_G \sim 0.013\ cm^{-1}{}^{\circ}C^{-1}$ which is in agreement with previous reported values.[50] Solving the heat diffusion equation, thermal conductivity of suspended CVD graphene was found to be $\sim 1800\ W/mK$. This value is in line with previous reported value of thermal conductivity of CVD graphene[56,60] and stands below the one for mechanically exfoliated graphene from HOPG.[53-55] The latter is due to the possible presence of grain boundaries and defects induced during synthesis and transfer process.

For the extraction of thermal conductivity from experimental data, Fourier's equation was solved in a 2D structure applying the specific geometry of suspended graphene over square heat sink. A numerical solution of this equation was conducted using COMSOL multi-physics package under corresponding boundary conditions. A Gaussian heat distribution of power was defined to model the laser induced heating:

$$P(x,y) = A \exp\left(-\frac{x^2+y^2}{2\sigma^2}\right) \qquad (1)$$

Where A is the amplitude of the heat source at the center of laser spot ($x = y = 0$) and is defined by setting the integral of $p(x,y)$ to the total absorbed power. The standard deviation of Gaussian power ($\sigma$) is calculated by setting the full width at half maximum (FWHM) of Gaussian power distribution to the experimental laser spot size ($0.36\ \mu m$). For this simulation an ideal heat sink was assumed with no thermal contact resistance between graphene and gold substrate. In order to extract thermal conductivity from the solutions of Fourier's equation, a reiterative process was followed. In this process, the laser power and thermal conductivity are given as input to the COMSOL model and the temperature profile is achieved as an output. By adjusting the temperature profile to the experimental measured value, thermal conductivity is iteratively extracted. The process is eased by defining the slope parameter: $\theta = \frac{\partial\omega}{\partial p} = \chi_G \frac{\partial T}{\partial p}$ and plotting the thermal conductivity versus $\theta$ (Fig 3(d)). Having the slope parameter from power dependent OTR measurement, one can easily extract the thermal conductivity.

**[Fig 3]**

The OTR measurement of graphene's thermal conductivity: (a) Scanning electron microscopy image of CVD-grown graphene transferred over gold TEM grid. The grid, shown in gold color, contains an array of $7.7\ \mu m$ square holes, shown in black. The transparent greenish area shows the suspended graphene flake covering holes fully and partially. (b) Raman spectrum of pristine CVD graphene suspended over TEM grid. (c) Linear shift of graphene's Raman G-peaks position with increasing excitation laser power. The slope of this plot, $\theta$, is later used to extract thermal conductivity. The inset shows scanning electron microscopy image of corresponding suspended flake. (d) The plot of thermal conductivity versus $\theta$ obtained from solving heat diffusion equation and used for extraction of thermal conductivity. Fig (a), (b) and (c) are adopted from Ref. 6 published by Royal Society of Chemistry.

## b) E-beam Irradiation of Graphene and Raman Study of the Induced Defects

In order to study the effect of defects on thermal conductivity of graphene, defects were induced controllably using low energy ( $20\,KeV$ ) electron beam irradiation.[89,90] Keeping the exposure area constant, the irradiation dose was managed by setting the beam current and exposure time. A Faraday cup was used to measure the beam current prior to irradiation process and the current values of $\sim 3\,nA$ to $\sim 10\,nA$ were applied. Controlled amount of defects were induced conducting multiple steps of irradiation. In addition to the use of Raman spectroscopy for thermal studies, the authors performed Raman analysis in two more capacities: quantifying the amount of induced defects and probing the nature of them. Therefore, Raman spectroscopy was conducted after each step of irradiation. The evolution Raman spectrum of CVD graphene after four steps of irradiation is shown in Fig 4(a). As more defects are induced, the D peak grows leading to the Raman D to G band intensity ratio ($I_D/I_G$) to enhance from its initial value ~ 0.13 all the way to ~ 1.00. As discussed in section 1, this intensity ratio is proportional to the square of inter defect distance ( $I_D/I_G \propto L_D{}^2$ ).[3,91] By curve fitting the theoretical expectation with experimental values, Ferrari et al.[20] introduced a set of empirical equations for extracting density of point defects having $I_D/I_G$ for a given excitation wavelength ($\lambda$). The equation is valid only for low defect density regime ($L_D < 10\ nm$) and is provided below:

$$N_D(cm^{-2}) = \frac{(1.8\ \pm 0.5) \times 10^{22}}{\lambda^4}\left(\frac{I_D}{I_G}\right) \qquad (2)$$

For low defect density regime, a linear increasing dependence of $I_D/I_G$ with irradiation dose is expected[89] and was confirmed for all irradiated graphene samples (Fig 4(b)). In order to probe the nature of defects using Raman spectroscopy, the procedure reported in Ref. 23 and discussed in section 1 was followed and vacancy type defects were found to be present in all irradiated graphene samples ($I_D/I_{D'} \sim 7$).

## c) Thermal Conductivity of the Irradiated Graphene with Defects

OTR measurements were performed after each step of irradiation in order to investigate the variation of the thermal conductivity with the density of induced defects. Fig 4(c) shows the results of calibration measurement and 4(d) power dependent OTR measurement for two levels of defect densities. It was found that the temperature coefficient of Raman G peak is not significantly influenced by the density of defects and therefore was assumed constant. The power dependent Raman measurement (fig 4(d)) however was reported to be challenging for highly irradiated samples. As the laser power is increased to locally heat the sample, the local sample temperature reaches the minimum value required for healing e-beam induced defects. The healing process was observed through the gradual reduction of the Raman D to G band intensity ratio and was avoided by keeping the laser power below the critical healing point ($\sim 2mW$). Fig 4(d) shows the power dependent OTR results in this reduced power range. A significant increase in the slope factor ($\theta$) was detected with increasing $N_D$, which is a direct sign of a suppressed thermal conductivity. Solving the Fourier's equation in a 2D geometry of the square graphene (section 3(a)) the thermal conductivities were extracted. The variation in thermal conductivity is plotted in Fig 4(e) as a function of its corresponding defect density ($N_D$). For small values of $N_D$, a linear decreasing regime of thermal conductivity was observed. The linear trend was then followed by a saturation behavior as density of defects exceeds $1.5 \times 10^{11} cm^{-2}$. It was found interestingly that the saturation of thermal conductivity occurs at relatively high $K$ values ($K \sim 400\ W/mK$), significantly higher than amorphous carbon limits.[55] In order to understand this intriguing behavior, the authors performed both a Boltzmann transport equation analysis and a molecular dynamics (MD) simulation.

**[Fig 4]**

The effect of electron beam irradiation on thermal conductivity of graphene: (a) The evolution of graphene's Raman spectrum with concentration of defects. As more steps of irradiation are applied, the Raman D and D′ bands evolves, leading to the Raman D to G band intensity ratio to grow from its initial value ~0.13 all the way to ~1. This ratio is later used to extract the density of defects. (b) The linear growth of the Raman D to G band intensity ratio with the irradiation dose. (c) The temperature dependence of the Raman G peak before and after e-beam irradiation, holding almost a same value of temperature coefficient. (d) The Raman G-peak shift with increasing excitation power recorded for the graphene subjected to two different amounts of irradiation. (e) The experimental (points) and theoretical (lines) variation in the thermal conductivity with the density of defects. The theoretical calculations were performed using BTE analysis and plotted for different specularity parameters. Fig (a), (b) and (e) are adopted from Ref. 6 published by Royal Society of Chemistry.

## d) The Dependence of Thermal Conductivity on the Defect Density

A Boltzmann transport equation (BTE) approach was applied within relaxation time approximation (RTA) to analyze the experimental data. The details of this analysis are provided in Ref. 6. Three scattering mechanisms of phonons were included in the BTE analysis: Umklapp scattering, boundary scattering and scattering on point defects. The total relaxation rate ($\tau_{tot}$) is defined as: $1/\tau_{tot} = 1/\tau_{Umklapp} + 1/\tau_{Boundary} + 1/\tau_{Point-Defect}$. The relaxation rate on boundaries ($\tau_{Boundary}$) is determined by introducing the specularity parameter ($p$) where $p$ represents the probability of specular scattering. A zero value of specularity parameter demonstrates a fully diffusive scattering regime with highest thermal resistance at the edge while $p = 1$ represents a specular scattering on the edge with no additional thermal resistance. The performed BTE analysis results are shown in Fig 4(e) along with the experimental data points. The results are plotted for varying $p$ values covering the experimental data points at a reasonable range ($0.5 < p < 0.9$).[71] The relaxation time on point defects ($\tau_{Point-Defect}$) is defined by introducing $\Gamma$ parameter, $\Gamma = \xi\ (N_D/N_G)$, representing the strength of point defect scattering. Here, $N_G$ denotes the concentration of carbon atoms and $\xi$ is the mass difference parameter depending on the mass of carbon atom ($M$) and the mass difference ($\Delta M$) between defects and carbon atom ($\xi = (\Delta M/M)^2$). For pure vacancy-type defects, which was found to be present in the irradiated samples via Raman analysis, $\xi$ is expected to be $\sim 9$ using perturbation theory calculations.[92] However an agreement between BTE analysis results and experimental data points was reached for much higher $\xi$ values ($\sim 590$). One should note that the sample could contain various types of defects including the initial defects existing in CVD graphene prior to irradiation. Moreover, not all types of defects can be probed by Raman spectroscopy and only defects that scatters electrons between the two valleys K and K′ of Brillouin zone can give rise to D-band.[44, 93,94] Nevertheless, the BTE analysis results confirm that an interplay of the three scattering mechanism could reproduce the experimental saturation behavior of $K\ (N_D)$.

In order to investigate the possible nature of defects caused by e-beam irradiation more in detail, one should consider that the minimum energy required to knockout carbon atoms and create vacancies in graphene is $\sim 80\ KeV$.[89,95-97] Therefore, the irradiation process under low energy ($20\ KeV$) electron beam could not by itself induce vacancies. However, the beam energy is sufficient to break carbon-carbon bonds and create functionalized epoxy and hydroxyl groups as a result of chemical reaction with $H_2O$ and $O_2$ molecules on graphene's surface. The temperature of graphene during OTR measurement ($\sim 350\ K$) provides enough energy for these functional groups to overcome the potential barrier (0.5-0.7 $eV$)[98] and diffuse. As more and more functional groups are created upon irradiation, these groups can come together and create vacancies by releasing $CO/CO_2$.[99] The formation of these functional groups could explain the higher $\xi$ value used in BTE analysis. Therefore, the strong suppression of thermal conductivity at low defect densities was explained by the formation of –C=O and other functionalized defects. As the population of these groups increases, continuous irradiation leads to the creation of single and double vacancies and saturates $K\ (N_D)$. It was confirmed by MD simulation that a combination of single and double vacancies could cause the saturation behavior observed experimentally.[6] The obtained results are useful for practical application of graphene in thermal management and for gaining better insight on phonon-point defect scattering in 2D materials.

# 4. Optothermal Raman Study of Graphene Laminate

In the previous sections we described the OTR technique in the context of measuring thermal properties of suspended 2D graphene layers. However, this technique is not limited to 2D materials and can be extended to 3D bulk samples, having sufficient laser power to induce local temperature rise. In this section we review the results obtained by Malekpour et al. using OTR method for thermal characterization of graphene laminate (GL) on polyethylene terephthalate (PET) substrate.[67] PET is a type of commercial plastic material used for manufacturing various containers. Being robust, lightweight and easy-to-work with, plastic is extensively used in packaging industry. However its weak thermal conduction limits its application where a conductive packaging material is desired. The wide use of plastic in electronic packaging has motivated the authors to study thermal properties of plastic covered with a thermal coating (graphene laminate). Graphene laminate is a type of graphene-based material consists of chemically synthesized graphene and FLG, conventionally used in coating applications.[100] In this type of coating the flakes are closely packed in overlapping structure. Coating of PET with a thin film of GL could compensate poor thermal properties of PET and lead to new application domains.

## a) Graphene Laminate as Thermal Coating

In this study[67] two sets of graphene laminate were prepared: as deposited and compressed, each containing three samples with varying mass densities ($1-1.9\ g/cm^3$) and thickness range ($9-44\ \mu m$). The laminates were produced by coating an aqueous dispersion of graphene on PET substrate via slit coater. Letting the films dry at low temperature (80°C), the as deposited GL-on-PET samples were prepared. A rolling compression was further applied to produce compressed samples. Scanning electron microscopy was performed to investigate the laminate's microstructure, both in top view (Fig 5(a)) and cross-sectional view (Fig 5(b)). Using cross-section SEM images the thickness of laminates were extracted (Table 1). As shown in top view SEM images, the laminates are made of stacking graphene and FLG with large size and shape variation orienting randomly. Most of the flakes are well aligned along PET substrate. However, small portion of the flakes are vertically oriented, observed in white bright areas under SEM. The rolling compression of the laminates substantially reduces the number of these misaligned flakes (Fig 5(a)).

Statistical data analysis was performed on the length of graphene flakes to define an average value for each individual sample. The average flake size is specifically important for the subsequent quantitative analysis of thermal conductivity. However, due to the large variation observed in both size and shape of the flakes, the task was challenging. Extensive top view SEM images were taken for this purpose. An average flake size was defined for each flake by taking the average of three different lengths (the inset in Fig 5(c)). For each sample, more than one hundred flakes were taken into account to increase the accuracy of the analysis. The behavior of the average flake size with the number of included flakes is shown in Fig 5(c). As the number of included flakes exceeds 50, the convergence of the average flake size to its apparent value was observed. The obtained values of the average flake size are provided in Table 1, varying from 0.96 to 1.24 $\mu m$.

**Fig [5]**

The OTR study performed on the macro-scale GL-on-PET samples: (a) Top view SEM image of the compressed and uncompressed laminates showing graphene flakes with arbitrary shape and alignment stacked in overlapping structure. While most of the flakes are oriented along PET substrate, some are misaligned appearing in bright regions. The population of misaligned flakes is significantly reduced after rolling compression. (b) Cross-sectional SEM image of GL-on-PET used for extracting the laminate thickness. The burgundy layer shows graphene laminate while the yellowish area demonstrates PET substrate. (c) Statistical data analysis performed on the length of graphene flakes in order to obtain an average value. The obtained average length is plotted versus the number of flakes, $N$, included in the study. As this number goes beyond $100$, the average length becomes independent of $N$. The inset shows how a single length is defined for a given flake of an arbitrary shape, averaging three different lengths. (d) The temperature dependence of Raman G-peak plotted for two GL-on-PET samples. (e) The shift over G-peak versus increasing excitation laser power. The two data sets with different slopes correspond to two laminates with varying thermal conductivities. Fig (b), (c), (d) and (e) are adopted from Ref. 67 published by American Chemical Society.

## b) Macro-Scale Configuration of the Optothermal Raman Measurement

OTR measurements were performed in macro-scale configuration. A special designed sample holder was used to suspend the GL-on-PET over two heat sinks. The designed sample holder along with the suspended sample is shown in Fig 2(d). The sample holder is made of a $2\,cm$ wide insulator trench with two massive aluminum pads, clamping the sample and serving as heat sinks. There are two major differences for performing OTR in 3D macro-scale compared to the case of 2D. First, higher level of laser power is needed (up to $10\,mW$ in case of GL) in order to provide sufficient energy to create local temperature rise ($\Delta T$) in the film. Second, the laser induced heating distribute in three dimensions, both along the plane and through the film thickness. Therefore, in order to extract thermal conductivity, heat diffusion equation need to be solved in three-dimensional geometry. Moreover, due to the large thickness of the laminates ($9-44\,\mu m$) the whole laser power is absorbed by the sample. All the other steps are similar to the 2D case and were discussed in previous sections (Sections 2 and 3(a)). Fig 5 shows the results of calibration measurement (d) and power dependent Raman measurement (e) for two different GL-on-PET samples. Here we will discuss the procedure used in Ref. 67 for solving Fourier's equation in 3D. In a simple approach one can assume a spherical Gaussian distribution of power inside the film:

$$P(x,y,z) = \frac{2P_{tot}}{\sigma^3\sqrt{(2\pi)^3}}\exp(-\frac{x^2+y^2+z^2}{2\sigma^2}) \quad . \qquad (3)$$

Here $P_{tot}$ is the total laser power and $\sigma$ is the standard deviation of Gaussian function obtained from the laser spot size. The two heat sinks, clamping the sample at its two ends, were modeled by holding the temperature of the sample at ambient temperature. All other boundaries were defined as insulator from environment by setting the temperature gradient across the boundary to zero ($\vec{n}.(K\nabla T) = 0$). The iterative $K$ extraction procedure, discussed in section 3(a), was followed in order to obtain the values of thermal conductivity. The values are provided in Table 1.

In order to more accurately model the laser induced heating, one can consider the laser penetration depth ($d$) in defining the power distribution function along z-direction (laminate thickness). In this case, $P(x, y, z)$ can be written as:

$$P(x, y, z) = \frac{P_{tot}}{2\pi d\sigma^2} \exp\left(-\frac{x^2 + y^2}{2\sigma^2}\right) \exp\left(-\frac{z}{d}\right). \qquad (4)$$

The power function in Eq. (4) is revised so that the total power along the sample volume is equal to the total absorbed power. The penetration depth could be obtained from the wavelength dependent refractive index of the graphene laminate.[101] Approximating it with the refractive index of graphite,[102] the $488\ nm$ laser penetration depth was found to be $\sim 30\ nm$ inside graphene laminate. Fig 6(a) shows a comparison between $K\ (\theta)$ plot obtained from the simple spherical model of $P(x, y, z)$ (Eq 3) and the more physically accurate model that includes penetration depth (Eq 4). The inclusion of penetration depth enhances the extracted thermal conductivity values by $\sim 10\%$.

The thermal conductivities of GL-on-PET samples are shown in Table 1 and plotted in Fig 6(b) versus their corresponding average flake length.[67] All the samples reveal high values of thermal conductivity, $\sim 40$ to $\sim 90$ ($W/mK$), which is much larger compared to thermal conductivity of PET by itself, $\sim 0.15$ to $\sim 0.24$ ($W/mK$). Therefore, by applying only a thin coating of graphene laminate, the thermal conductivity of plastic was enhanced up to $\times 600$ times. The high values of thermal conductivity was achieved both in compressed and uncompressed samples suggesting that one could benefit form GL coating even without any additional processing. No correlation was found between thermal conductivities of the laminates with neither their thicknesses nor their mass densities. However, it was found that the thermal conductivity of GL-on-PET scales up linearly as the average length of the flakes increases both in compressed and uncompressed samples (Fig 6(b)). The values of thermal conductivity for compressed samples stands above those of uncompressed owing to the better flakes attachment and coupling (Fig 5(a)). The linear correlation of thermal conductivity with the average flake length suggests that the phonon transport in graphene laminate is dominated by scattering from flake boundaries rather than intrinsic properties of individual flakes. To gain better insight onto the length dependence of thermal conductivity, theoretical calculations were performed.

Table 1. Physical and thermal characteristics of the graphene laminate samples

**[Fig 6]**

Thermal conductivity of GL-on-PET: (a) Thermal conductivity extraction plot, $K\ (\theta)$, considering two different models of laser heating, spherical heat source that assumes a symmetric distribution in $XYZ$ directions and the more accurate model which includes laser penetration depth in $Z$direction. The inclusion of the penetration depth enhances the extracted $K$ values by $\sim 10\%$. The inset shows temperature profile along laminate thickness considering spherical heat source model. (b) The dependence of thermal conductivity on the average flake length plotted for both uncompressed (blue rectangles) and compressed (red circles) GL-on-PET. The thermal conductivity scales up linearly with the average flake length. While the both types of laminates show high values of thermal conductivity, compressed laminates provide better heat conduction properties owing to the improved flake alignment as a result of compression. (c) Thermal conductivity of graphene laminate obtained from theoretical calculations as a function of temperature. The results are plotted for different flake length $D$ and defect scattering strength $\Gamma$ along with the experimental data points for comparison. The figure demonstrates a weak temperature dependence of GL-on-PET in the range of experimental data points. (d) The calculated temperature dependence of thermal conductivity plotted for large flake length $D$ and small $\Gamma$ parameter. In this regime, the Umklapp limited temperature behavior of thermal conductivity observed in crystalline graphene is restored. Fig (b), (c) and (d) are adopted from Ref. 67 published by American Chemical Society.

## c) Theoretical Analysis of the Size Dependence of Thermal Conductivity

Thermal conduction along the in plane direction of GL-on-PET could be modeled considering the heat conduction in each individual flake as well as their interfacial thermal resistance. The latter is defined by the strength of flakes attachment and their mutual orientation. However, Do to the uncertainty of such parameters, defining a thorough model that includes all these parameters would be challenging. The author introduced a simple model to analyze the length dependence of thermal conductivity in an individual flake of few layer graphene (FLG). The task was performed by modifying the formula derived for heat conduction in thin films of graphite.[103,104] The characteristics of graphene laminate are entered to the formula considering the average length of the flakes as well as their defect concentrations. The in-plane thermal conductivity of graphene laminate can be written as:[103,105,106]

$$K = K_{xx} = \frac{1}{L_x L_y L_z} \sum_{s,\vec{q}} \hbar\omega_s(\vec{q})\tau(\omega_s(\vec{q}))\, v_{s,x} v_{s,x} \frac{\partial N_0}{\partial T} \qquad (5)$$

Where $\tau\ (\omega_s(\vec{q}))$ shows the relaxation time of phonons having frequency of $\omega_s(\vec{q})$. The summation is performed over all acoustic phonon branches: longitudinal acoustic (LA), transverse acoustic (TA) and out of plane acoustic (ZA) mode. Here $\vec{q}(q_{\|}, q_z)$ is the phonon wave vector, $T$ denotes the temperature, $L_x$, $L_y$, $L_z$ are the sample dimensions and $v_{x,s}$ is the group velocity of phonons along branch direction. The phonon transport in graphene laminate is treated as 2D and 3D regimes based on the phonon frequencies. Defining $\omega_c$ as a certain low band cut off frequency, phonons with higher frequencies are transported as 2D gas while those with frequencies below $\omega_c$ are treated as 3D.[103] Following the approximation procedure used in Ref. 103 the 2D and 3D parts of thermal conductivity can be derived as:

$$K_{3D} = \frac{\hbar^2}{4\pi^2 K_B T^2} \sum_{s=LA,TA,ZA} \frac{1}{v_s^{\perp}} \int_0^{\omega_{c,s}} [\omega_s^{\|}(q^{\|})]^3 \tau(\omega_s^{\|}) v_s^{\|}(q^{\|}) \frac{\exp\left(\frac{\hbar\omega_s^{\|}}{K_B T}\right)}{\left[\exp\left(\frac{\hbar\omega_s^{\|}}{K_B T}\right) - 1\right]^2} q^{\|} d\omega_s^{\|}$$

$$K_{2D} = \frac{\hbar^2}{4\pi^2 K_B T^2} \sum_{s=LA,TA,ZA} \frac{\omega_{c,s}}{v_s^{\perp}} \int_{\omega_{c,s}}^{\omega_{max,s}} [\omega_s^{\|}(q^{\|})]^2 \tau(\omega_s^{\|}) v_s^{\|}(q^{\|}) \frac{\exp(\frac{\hbar\omega_s^{\|}}{K_B T})}{[\exp\left(\frac{\hbar\omega_s^{\|}}{K_B T}\right) - 1]^2} q^{\|} d\omega_s^{\|}$$

(6)

The total thermal conductivity can be achieved adding the two components. The cut off frequency is branch dependent and is defined as the frequency of phonons at A-point of Brillouin zone and $v_s^{\perp} = \omega_{c,s}/q_{z,max}$. Three phonon scattering mechanisms were included in the model:[104-107] Umklapp scattering, point defect scattering and scattering on the boundaries of graphene flakes with corresponding relaxation times ($\tau$):

$$\tau_{Umklapp}(\omega_s^{\|}) = M v_s^2 \omega_{max,s} / \gamma_s^2 K_B T (\omega_s^{\|})^2$$

$$\tau_{Point-Defect}(\omega_s^{\|}) = 4 v_s^{\|} / S_0 \Gamma q (\omega_s^{\|})^2$$

$$\tau_{Boundary}(\omega_s^{\|}) = D / v_s^{\|}$$

(7)

Where $\gamma$ is the average Gruneisen parameter defined for each branch, $\omega_{max,s}$ is the branch dependent maximum frequency, $S$ is the cross section area per atom, $M$ is the mass of graphene's unit cell and $\Gamma$ is a parameter showing the strength of point defect scattering obtained form the density of defects as well as their mass densities. Based on Matthiessen's rule the total relaxation time is: $1/\tau_{tot} = 1/\tau_{Umklapp} + 1/\tau_{Boundary} + 1/\tau_{Point-Defect}$. A fully diffusive boundary scattering was assumed in the model by setting the specularity parameter to zero. Performing energy dispersive spectroscopy (EDS) on GL-on-PET, $\Gamma$ parameter was estimated. Fig 6(c) shows the results of these calculations for varying flake length and $\Gamma$ values along with experimental data points (red and pink circles). One should notice that, for the experimental range of $\Gamma$ and $D$ parameters, the thermal conductivity of graphene laminate is not significantly influenced by its temperature. This behavior resembles the characteristics of polycrystalline materials were the dominant scattering mechanism is on grain boundaries.[107] The expected temperature dependence of thermal conductivity for crystalline graphene[108] is restored as $\Gamma$ moves toward zero defect regimes and the flake size increase to $30\ \mu m$ (Fig 6(d)). It was confirmed theoretically that an increase in the flake length leads to an enhancement in the thermal conductivity of graphene laminate. However, the length dependence of thermal conductivity was found to be weaker than that of experimentally observed. This was attributed to parameters such as flake orientation and coupling that were not included in the model.

Ref. 109 introduces a more accurate model for thermal conductivity of graphene laminate, which includes the effect of flakes coupling, by performing MD simulations. The reported results[109] confirm the linear length dependence of thermal conductivity in graphene laminate. This length dependence is also in agreement with previous literature reports on carbon nanotubes and other carbon allotropes.[110,111] Despite the significant improvement in thermal conductivity of GL-on-PET compared to PET, the $K$ values are still substantially lower than that of graphene.[55] However one should note that phonon transport in graphene laminate is dominated by flake length, orientation and coupling rather than intrinsic properties of graphene. The study of thermal conduction in GL-on-PET is helpful for the use of plastic in new application domains such as electronic device packaging where dissipating the excess heat from the device is critical.

# 5. Optothermal Raman Study of Other 2D Materials

In the previous sections, we reviewed OTR investigation of the thermal properties of graphene-based films both in 2D monolayer of graphene and 3D macro-scale graphene laminate. The micro-Raman thermometry is not limited to graphene-based films and can be extended to other 2D materials with clear Raman signature.[62-66] Despite the excellent electrical and thermal properties of graphene, the lack of band gap limits its application in device fabrication. Two-dimensional transition metal dichalcogenides (TMDs) has recently attracted lots of interest due to their superior properties.[112-114] The presence of band gap in TMDs make them more favorable for device applications. Understanding thermal properties of these materials is crucial for the development of TMD based electronic devices where thermal management is critical. OTR analysis has been widely used to study thermal properties of two-dimensional TMD materials.[62-66] Here in this section we will review some of the recent OTR studies performed on 2D lattice of $MoS_2$, $WS_2$ and $MoSe_2$ in mono and bilayer structure.[62,64,65]

## a) Raman Study of Thickness in Transition Metal Dichalcogenides

Similar to graphene, Raman spectroscopy could be used to investigate the number of layers in few layer TMDs. It is believed that the number of layers strongly correlates with van der waals interactions between the layers. These interactions influence both the band structures[115,116] and the lattice vibrations.[117] It is reported that the wavenumber difference between two prominent Raman bands, the in-plane $E_{2g}^{1}$ and the out of plane $A_{1g}$ modes, is used to measure the number of layers in $MoS_2$ and $WS_2$ TMDs.[64,117] An increase in the number of layers leads to the out of plane $A_{1g}$ mode to blue shift owing the enhancement of restoring forces. At the same time the in-plane $E_{2g}^{1}$ mode red shifts as the number of layers increases, attributed to Coulombic forces and the effect of additional interlayer interactions as a result of stacking.[117,118] Therefore, by monitoring the difference between wavenumbers of the two Raman peaks, one can conveniently determine the thickness with atomic-level accuracy. Ref. 117 reported the first use of Raman spectroscopy as thickness indicator in few layer $MoS_2$. For this study mechanically exfoliated one to six layers of $MoS_2$ were prepared on $SiO_2$/Si substrate and their Raman spectra were studied. Atomic force microscopy was used to independently determine the number of atomic layers (Fig 7(a)). The variation of Raman spectra with the film thickness is plotted in Fig 7(b) showing the stiffening of out of plane $A_{1g}$ mode and softening of in-plane $E_{2g}^{1}$ band. It was found that the variation in the wavenumber of $A_{1g}$ mode is twice larger than $E_{2g}^{1}$ vibrational mode. Raman spatial mapping was further used to investigate the shift in the wavenumbers of these two modes (Fig 7(c,d)). For a given thickness, the $A_{1g}$ and $E_{2g}^{1}$ Raman peak frequencies of different locations shows a very small variation enabling an accurate measurement of film thickness using the two peak frequencies. Moreover, the opposite variation of the $A_{1g}$ and $E_{2g}^{1}$ frequencies reduces the special sensitivity of this method. In addition to Raman spectroscopy, photoluminescence (PL) study could be used to confirm the presence of monolayer TMDs.[64,65] This is owing to the interesting feature of TMDs in transforming from indirect to direct band gap material as the thickness is reduced to monolayer.[119-122]

**[Fig 7]**

The Raman-based thickness detection in few layer $MoS_2$ (a) Atomic Force Microscopy image of mechanically exfoliated FL-$MoS_2$ on $SiO_2$/Si substrate showing regions of one to four layers. (b) The variation of FL-$MoS_2$ Raman spectra as the number of layers increases from one to six. As the number of atomic layers increases, the out of plane $A_{1g}$ mode blue shifts while the in-plane $E_{2g}^{1}$ red shifts. The wavenumber difference between two peaks could be used as thickness indicator. Raman spectrum of bulk $MoS_2$ is added for comparison. (c,d) Special Raman mapping using the frequencies of $E_{2g}^{1}$ (c) and $A_{1g}$ (d) modes. Figures are adopted from Ref. 117 published by American Chemical Society.

## b) Raman Peak Selection for Thermal Studies

Among TMDs, $MoS_2$ is one of the most stable ones, which attracted the highest attention. The first OTR study was performed by Sahoo et al. on few layer MoS2 films grown from vapor phase.[63] Yan et al. later used OTR technique to investigate thermal conduction in monolayer of mechanically exfoliated $MoS_2$.[62] Due to the weak van der waals forces, TMDs are easily exfoliated to atomically thin layers, similar to graphene.[123-127] However, CVD growth of TMDs is widely used for the synthesis of large single domain sizes.[128-131] The Raman spectrum of $MoS_2$ shows two distinct peaks: the in-plane $E_{2g}^1$ mode ($\sim 385\ cm^{-1}$) and the out of plane $A_{1g}$ mode ($\sim 405\ cm^{-1}$) for excitation wavelength of $514.5\ nm$.[117] The $E_{2g}^1$ vibrational mode corresponds to the in-plane motion of Sulfur and Molybdenum atoms in opposite directions while the $A_{1g}$ out of plane mode originates from the relative motion of Sulfur atoms. Fig 8(a) shows the Raman signal of monolayer $MoS_2$ in the wavenumber range of $E_{2g}^1$ and $A_{1g}$ mode, along with their corresponding motions.[62] Both these peaks are temperature dependent owing to the lattice thermal expansion which itself originate from anharmonicity of interatomic potentials. The two peaks are linearly softens as temperature increases. A simple lorentzian peak fitting is used on each mode to investigate their linear temperature coefficient ($\Delta\omega = -\chi_T \Delta T$).[62] One should note that the nonlinear temperature coefficients are usually observed in high temperatures and could be neglected.[53,63]

Ref. 62 reported the temperature coefficient of suspended $MoS_2$ to be $\sim 0.011\ cm^{-1}K^{-1}$ and $\sim 0.013\ cm^{-1}K^{-1}$ for $E_{2g}^1$ and $A_{1g}$ modes, respectively. Comparing the results with sapphire supported $MoS_2$, the author found a slight change in the temperature coefficient of $E_{2g}^1$ as a result of substrate induced strain, while $A_{1g}$ mode stays unaffected. Despite the strain sensitivity of $E_{2g}^1$ vibrational mode, both peaks can be used as a mean of thermometry. However, in case of $WS_2$, due to a strong overlap of $E_{2g}^1$ peak with other vibrational modes,[64] multiple Lorentzian fitting is required to extract the wavenumber of this peak, increasing the measurement error. Therefor $A_{1g}$ mode is chosen for OTR studies.[64] Fig 8(b) shows the Raman signature of mono and bilayer $WS_2$ film for the wavenumber ranging from 300 to $600\ cm^{-1}$. The figure indicates the multiple curve fittings required for extracting $E_{2g}^1$ mode wavenumber and also validates our previous discussion on the thickness dependence of $E_{2g}^1$ and $A_{1g}$ modes.

## c) Optothermal Raman Study of Transition Metal Dichalcogenides

Though the substrate might not affect the temperature coefficient of Raman peaks, it could play an important role in heat dissipation. Therefore suspended samples are used for OTR studies. Yan et al. performed OTR study on mechanically exfoliated monolayer of $MoS_2$ over $1.2\ \mu m$ wide holes of $Si_3N_4$ grid (Fig 8(c)).[62] The laser spot diameter (2R) was estimated to be $\sim 0.34\ \mu m$ considering laser wavelength ($\lambda$) and numerical aperture (NA) of the objective lens ($R = \lambda/\pi NA$). The authors claimed that due to a weak thermal conduction in $MoS_2$, small hole diameters ($1.2\ \mu m$) would still allow an accurate measurement of thermal conductivity. For the same reason, very small laser powers were implemented, below$0.25\ mW$, to avoid any possible laser damage. For the extraction of thermal conductivity, the absorption coefficient of ~9% was assumed for $488\ nm$ excitation.[62] The slope of power dependent Raman for $E^1_{2G}$ and $A_{1g}$ modes were measured to be around the same range, $-12.8\ cm^{-1}mW^{-1}$ and $-10.9\ cm^{-1}mW^{-1}$, respectively.

Peimyoo et al. studied thermal properties of CVD-grown single and bilayer WS2 suspended over $6\ \mu m$ holes on $SiO_2$/Si substrate using OTR technique.[64] Though both $E^1_{2g}$ and $A_{1g}$ Raman modes show excellent linear temperature dependences, the $A_{1g}$ mode was chosen for thermal conductivity extraction due to its single Lorentzian peak fitting. Fig 8(d) shows the power dependent Raman results obtained from monolayer suspended $WS_2$ along with cross sectional schematic of the experimental set up.[64] The literature reported value of absorption coefficient, 4% per layer for $\lambda = 532\ nm$, was used for the extraction of thermal conductivity.[132] The laser spot size was directly measured by line scanning profile of Raman signal across the sample edge ($\sim 1\mu m$). The large diameter of the holes minimizes any possible effect from sidewalls.

Cai et al. suggested the use of two different laser spot size in OTR study of $SiO_2$-supported and suspended graphene in order to extract the interfacial thermal conductance of $SiO_2$-graphene.[56] Following the idea, Zhang et al[65] performed OTR technique on monolayer and bilayer films of $MoS_2$ and $MoSe_2$ prepared by mechanical exfoliation. For this purpose, suspended MoS2 and MoSe2 films were prepared on gold and SiO2 substrates respectively. By comparing the response of fully supported and suspended samples using different laser spot diameters, the authors extracted the thermal conductivities of $MoS_2$ and $MoSe_2$ films as well as their interfacial thermal conductance with the substrate. The absorption coefficients were extracted by performing direct measurements on exfoliated films on quartz substrate. The details of these measurements are provided in Table 2.

**[Fig 8]**

Thermal conductivity of TMDs using OTR technique: (a) Raman spectra of monolayer $MoS_2$ in the wavenumber range of $E^1_{2g}$ and $A_{1g}$ modes collected at different temperatures. Dashed lines are added to show the peak shifts along with their corresponding motions. (b) Raman spectra of mono and bilayer $WS_2$ film plotted for the wavenumber ranging from 300 to 600 $cm^{-1}$. Unlike $MoS_2$, multiple peak fitting is required to extract $E^1_{2g}$ band wavenumber. (c) Schematic of experimental setup used for performing OTR study on monolayer $MoS_2$ on $Si_3N_4/SiO_2$/Si substrate. The mechanically exfoliated $MoS_2$ is suspended using $1.2\ \mu m$-wide holes of $Si_3N_4$. (d) The shift of out of plane $A_{1g}$ mode as a function of increasing excitation laser power recorded for monolayer suspended $WS_2$. The inset shows cross sectional schematic of the experimental set-up. Fig (a), (c) are adopted from Ref. 62 published by American Chemical Society and Fig (b), (d) are adopted from Ref 64 published by Springer.

## d) Extraction of Thermal Conductivity

Previously for the extraction of thermal conductivity of graphene on gold substrate[6] the gold heat sink was assumed to be ideal with zero contact resistance with graphene film. However, in order to be more accurate, one should consider the interfacial thermal resistance and other boundary conditions outside the suspended area. In Refs. 62,64,65 the authors solved the heat diffusion equation both inside and outside suspended area for extracting thermal conductivity of TMDs. Considering R to be the hole radius and using cylindrical coordinates, the heat distribution inside and outside the hole can be written as:

$$\begin{cases} K\frac{1}{r}\frac{d}{dr}\left[r\frac{dT_1(r)}{dr}\right]+q(r)=0 & r<R \\ K'\frac{1}{r}\frac{d}{dr}\left[r\frac{dT_1(r)}{dr}\right]-\frac{G}{t}\left[T_2(r)-T_a\right]=0 & r>R \end{cases} \qquad (8)$$

Where r is the radial position from the center of hole, $T_1(r)$ and $T_2(r)$ are the temperature distribution inside and outside the hole, respectively. $T_a$ is the ambient temperature, $K$ and $K'$ are thermal conductivity of suspended and supported TMD film which is usually assumed to be the same. $G$ denotes the interfacial thermal conductance between TMD and substrate, t is the film thickness and $q(r)$ is the volumetric Gaussian laser heating similar to Eq 1.[56,62,64] One should note that the heat transport equation outside the hole was written considering the heat transport along the supported TMD and the heat transport from TMD to the heat sink, which keeps the ambient temperature. Refs. 62,64 solve the two equations considering a typical value of interfacial thermal conductance $G$, while Ref 65 extracted the $G$ values performing measurements with two different spot size. For extracting the thermal conductivity, the average temperature rise inside the laser spot is compared to the experiment and $K$ values are iteratively extracted. Table 2 summarizes the OTR experimental details and thermal conductivities on graphene and other 2D materials reviewed in this paper. The knowledge of thermal conductivity of 2D materials could contribute to the growth of their application in electronic devices.

Table 2. The details of OTR studies performed on graphene and other 2D materials

Table 1

| GL-on-PET | Laminate Thickness [µm] | Average Flake Length [µm] | Thermal Conductivity [$Wm^{-1}K^{-1}$] | Laminate Type |
|---|---|---|---|---|
| 1 | 44 | 1.10 | 40.0±7.5 | Uncompressed |
| 2 | 14 | 1.15 | 59.0±3.6 | Uncompressed |
| 3 | 13 | 1.24 | 75.5±11.3 | Uncompressed |
| 4 | 9 | 1.18 | 90.0±9.4 | Compressed |
| 5 | 24 | 1.07 | 63.5±4.0 | Compressed |
| 6 | 30 | 0.96 | 44.5±6.9 | Compressed |

Table 2

| **Material** | **Substrate** | **Preparation** | **Absorption coefficient [%]** | **Laser wavelength [nm]** | **Hole size [µm]** | **Laser spot size [µm]** | $\chi_T$ **[$cm^{-1}K^{-1}$]** | **K [$Wm^{-1}K^{-1}$]** | **Reference** |
|---|---|---|---|---|---|---|---|---|---|
| **1L-$MoS_2$** | $Si_3N_4$ | Exfoliation | 9 | 488 | 1.2 | 0.34 | $E_{2g}^1$: 0.011<br>$A_{1g}$: 0.013 | 34.5 | [62] |
| **1L-$MoS_2$** | Au | Exfoliation | 5.8 | 532 | 4.0 | 0.46<br>0.62 | $A_{1g}$: 0.020 | 84 | [65] |
| **2L-$MoS_2$** | Au | Exfoliation | 12.1 | 532 | 3.0 | 0.46<br>0.62 | $A_{1g}$: 0.014 | 77 | [65] |
| **FL-$MoS_2$** | Cu | Vapor phase growth | - | 532 | - | 1-1.5 | $E_{2g}^1$: 0.013<br>$A_{1g}$: 0.012 | 52 | [63] |
| **1L-$WS_2$** | $SiO_2$ | CVD | 4 | 532 | 6.0 | 0.5 | $E_{2g}^1$: 0.012<br>$A_{1g}$: 0.014 | 32 | [64] |
| **2L-$WS_2$** | $SiO_2$ | CVD | 8 | 532 | 6.0 | 0.5 | $E_{2g}^1$: 0.013<br>$A_{1g}$: 0.011 | 53 | [64] |
| **1L-$MoSe_2$** | $SiO_2$ | Exfoliation | 5.6 | 633 | 2.5 | 0.46<br>0.62 | $A_{1g}$: 0.014 | 59 | [65] |
| **2L-$MoSe_2$** | $SiO_2$ | Exfoliation | 9.4 | 633 | 3.0 | 0.46<br>0.62 | $A_{1g}$: 0.009 | 42 | [65] |
| **SLG** | Au | CVD | 5.7 | 488 | 7.5 | 0.36 | $\chi_G$: 0.015 | 1800 | [6] |
| **SLG** | $SiO_2$ | Exfoliation | 13 | 488 | 3 | 0.5-1 | $\chi_G$: 0.016 | 4840-5300 | [53] |
| **SLG** | Au | CVD | 3.3 | 532 | 3.8 | 0.38<br>0.48 | - | 2500 | [56] |

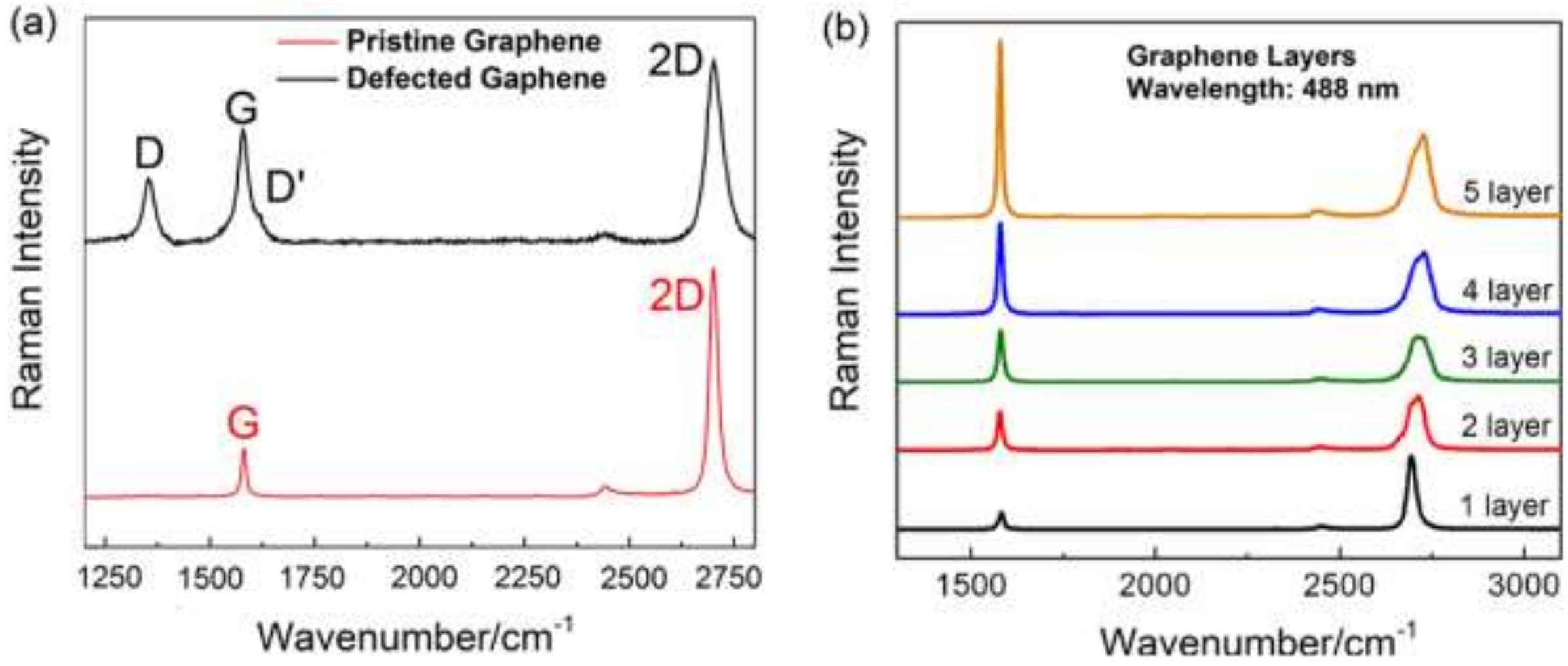

Fig 1

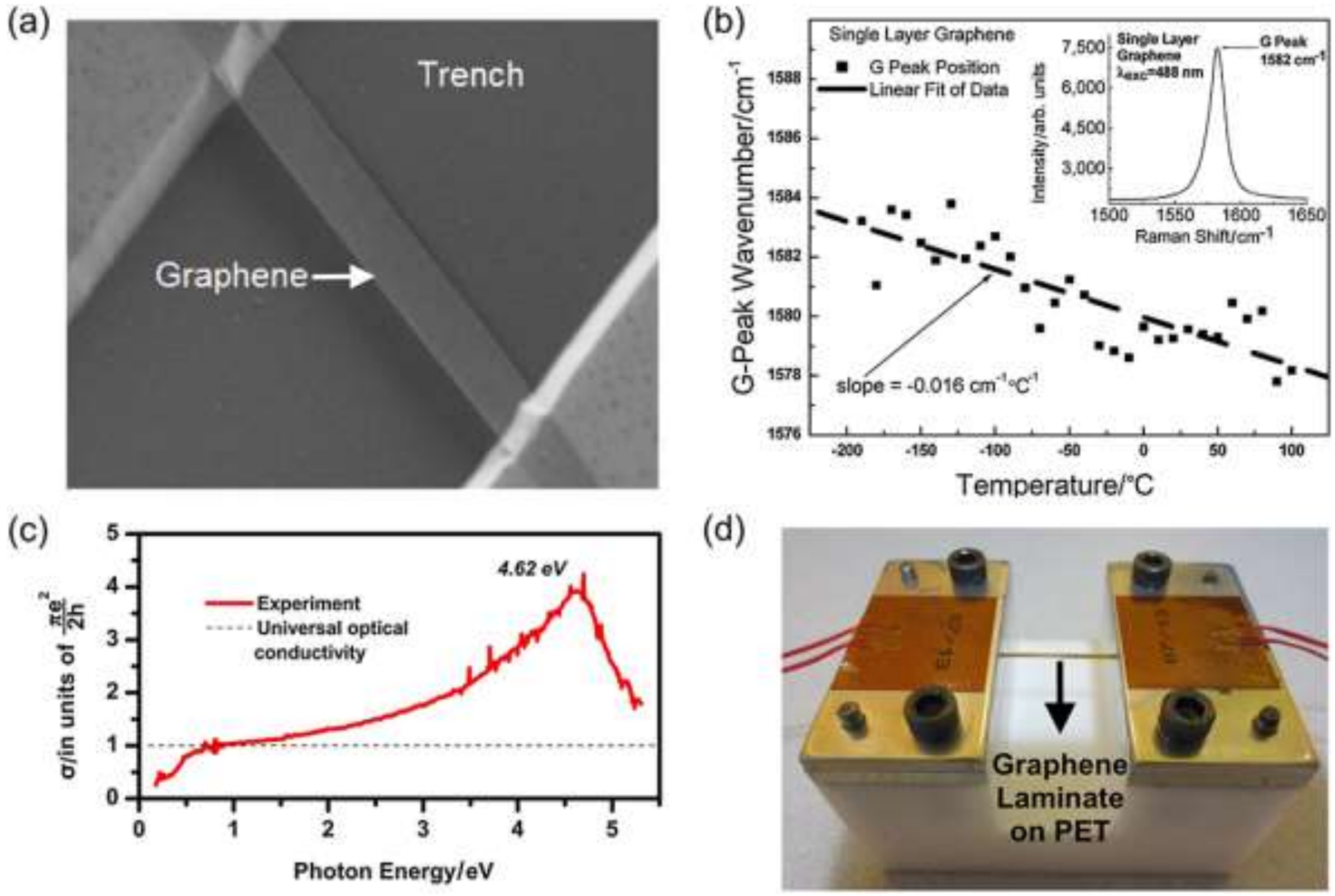

Fig 2

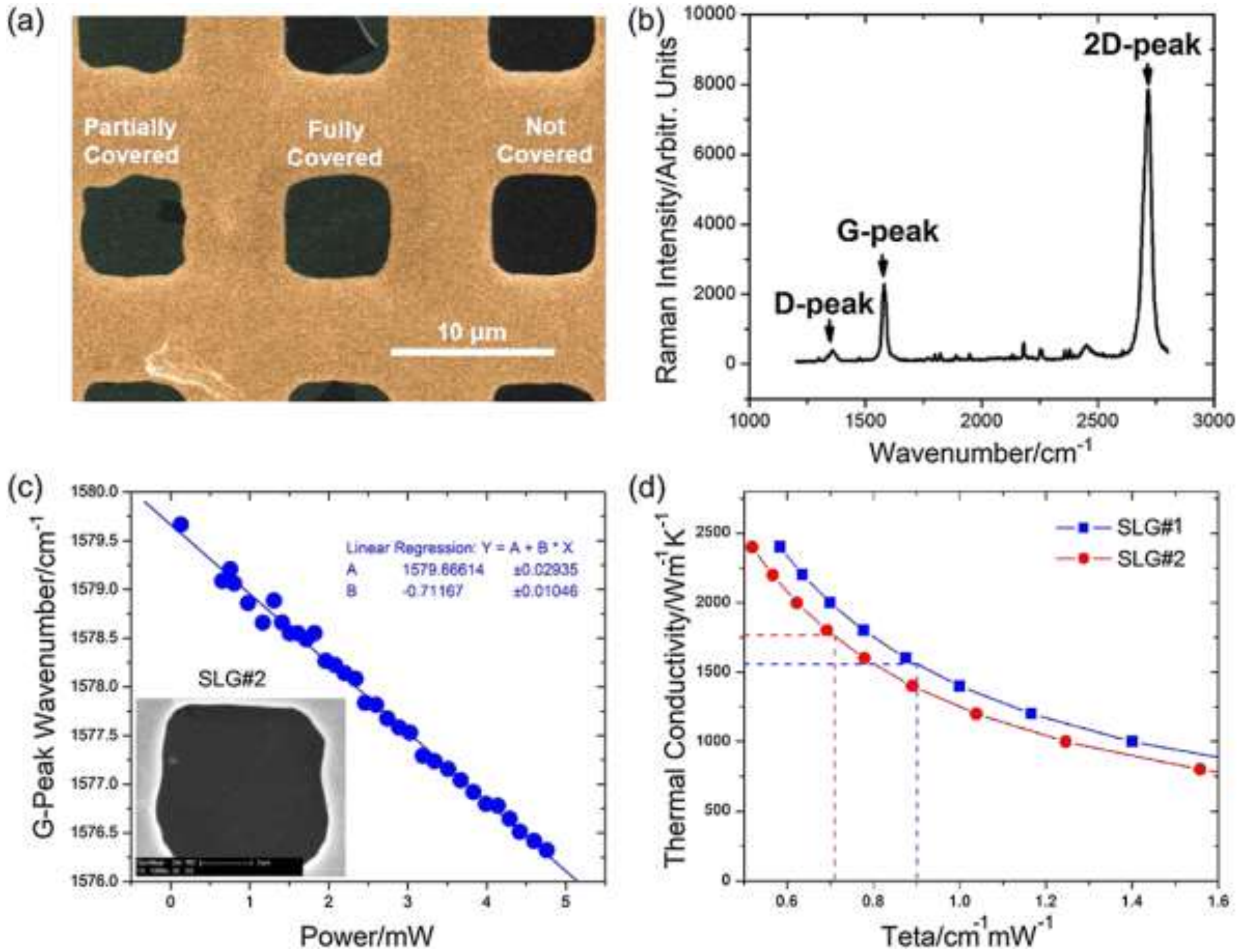
(a)
Partially Covered
Fully Covered
Not Covered
10 μm
(b)
Raman Intensity/Arbitr. Units
Wavenumber/cm-1
D-peak
G-peak
2D-peak
(c)
G-Peak Wavenumber/cm-1
Power/mW
Linear Regression: Y = A + B * X
A 1579.66614 ±0.02935
B -0.71167 ±0.01046
SLG#2
(d)
Thermal Conductivity/Wm-1K-1
Teta/cm-1mW-1
SLG#1
SLG#2

Fig 3

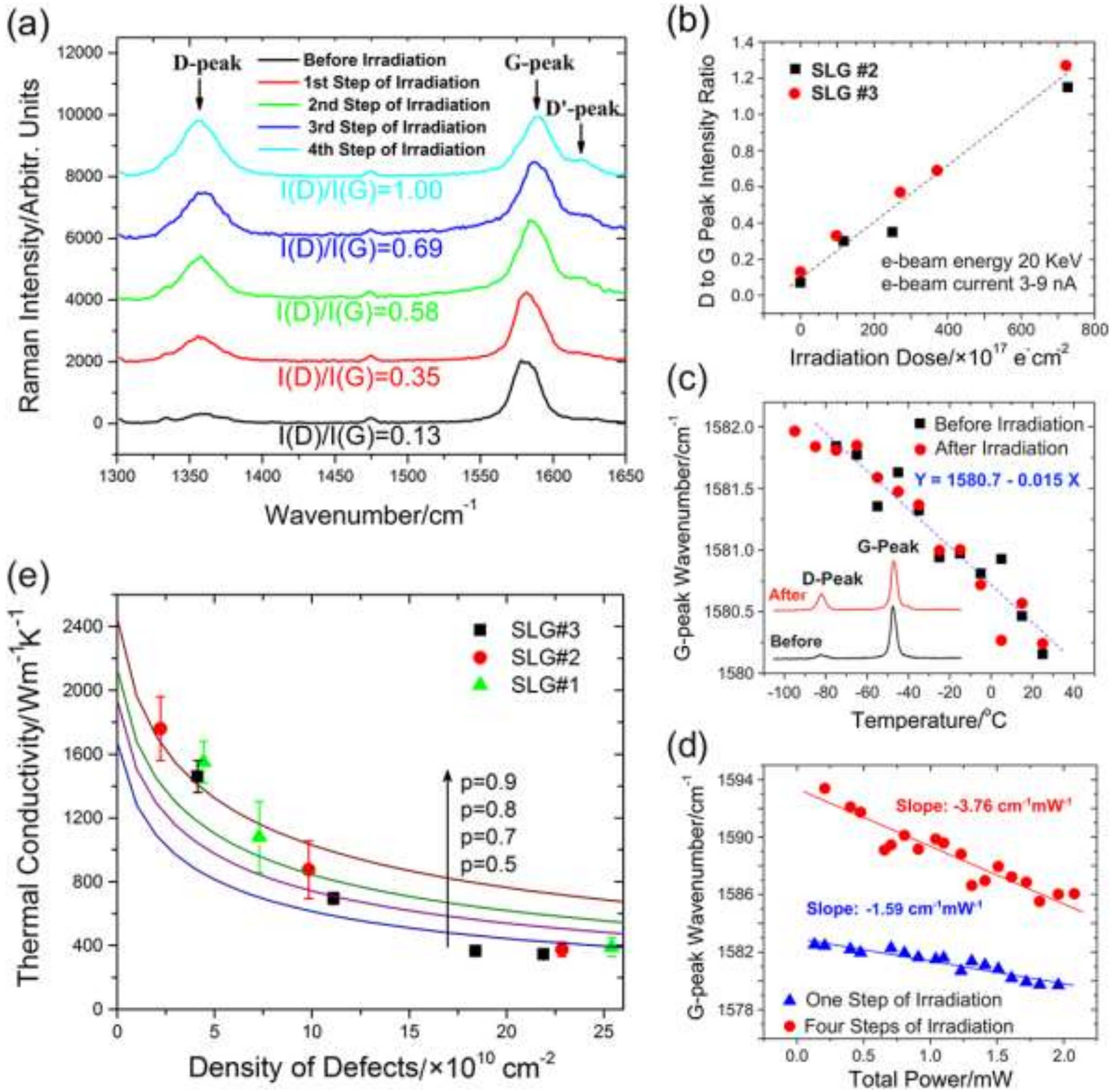
(a)
D-peak
G-peak
D'-peak
Before Irradiation
1st Step of Irradiation
2nd Step of Irradiation
3rd Step of Irradiation
4th Step of Irradiation
I(D)/I(G)=1.00
I(D)/I(G)=0.69
I(D)/I(G)=0.58
I(D)/I(G)=0.35
I(D)/I(G)=0.13
Raman Intensity/Arbitr. Units
Wavenumber/cm-1
(b)
SLG #2
SLG #3
e-beam energy 20 KeV
e-beam current 3-9 nA
D to G Peak Intensity Ratio
Irradiation Dose/×10^17 e-cm-2
(c)
Before Irradiation
After Irradiation
Y = 1580.7 - 0.015 X
G-Peak
D-Peak
After
Before
G-peak Wavenumber/cm-1
Temperature/°C
(d)
Slope: -3.76 cm-1mW-1
Slope: -1.59 cm-1mW-1
One Step of Irradiation
Four Steps of Irradiation
G-peak Wavenumber/cm-1
Total Power/mW
(e)
SLG#3
SLG#2
SLG#1
p=0.9
p=0.8
p=0.7
p=0.5
Thermal Conductivity/Wm-1K-1
Density of Defects/×10^10 cm-2

Fig 4

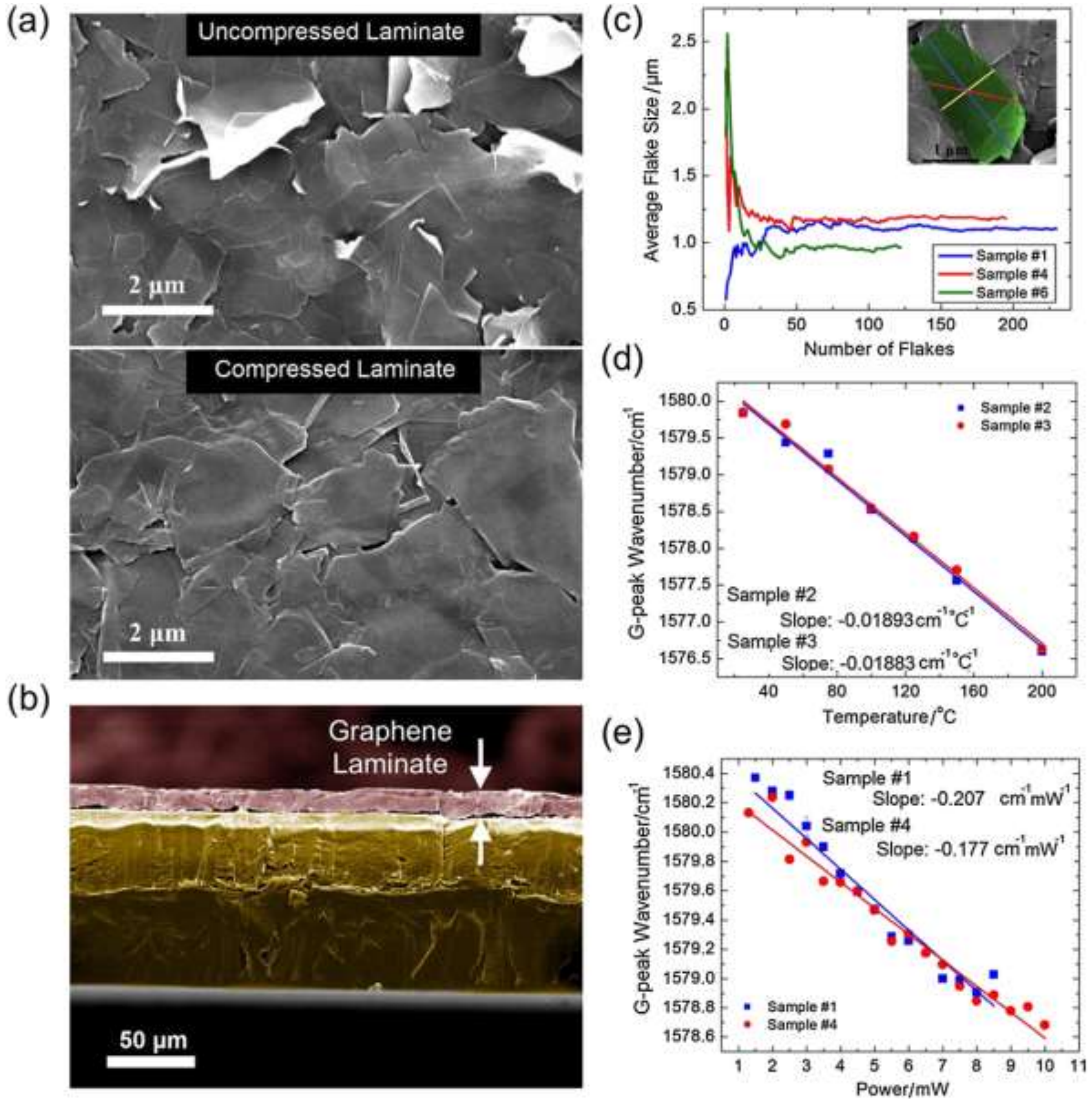

Fig 5

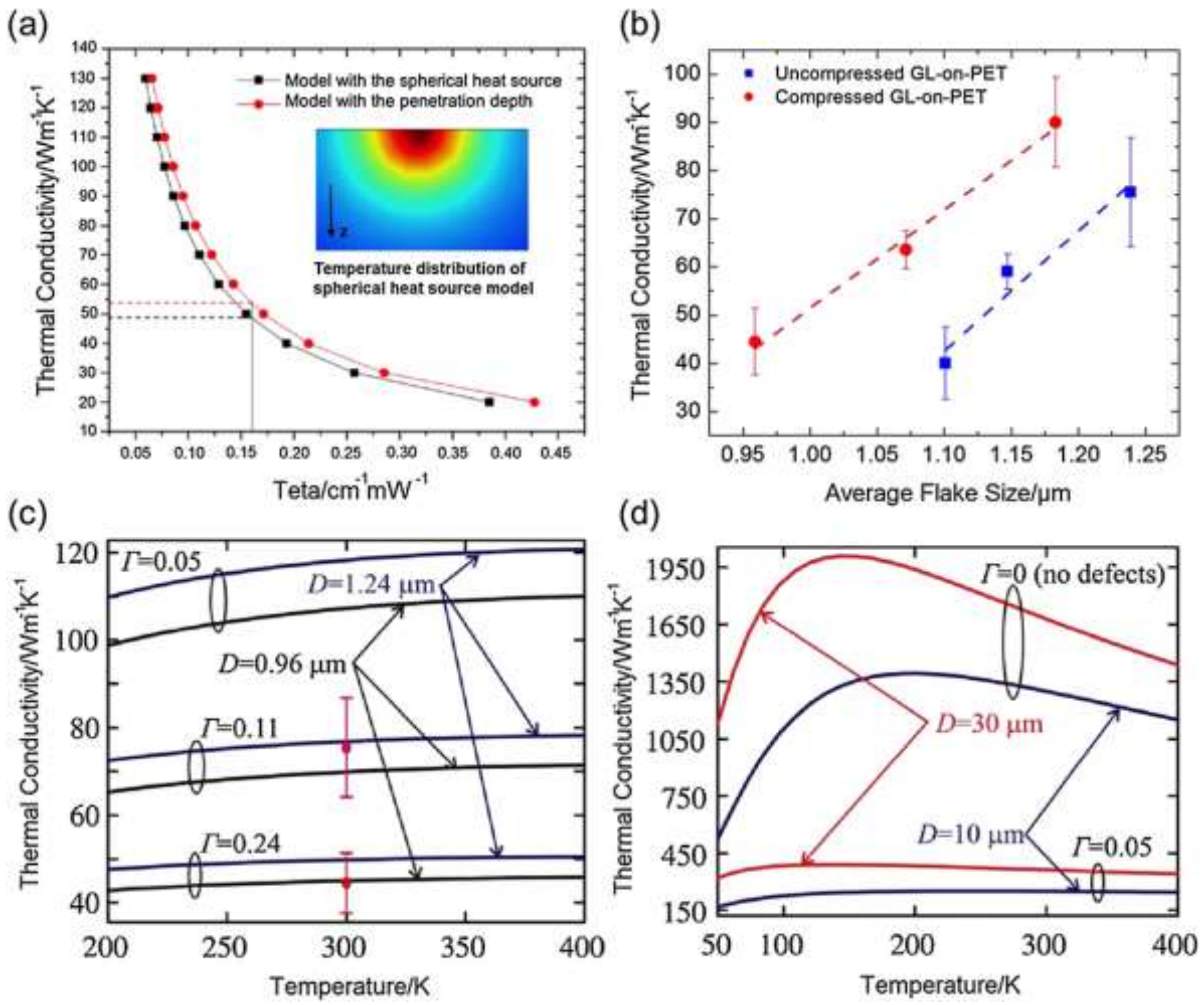

Fig 6

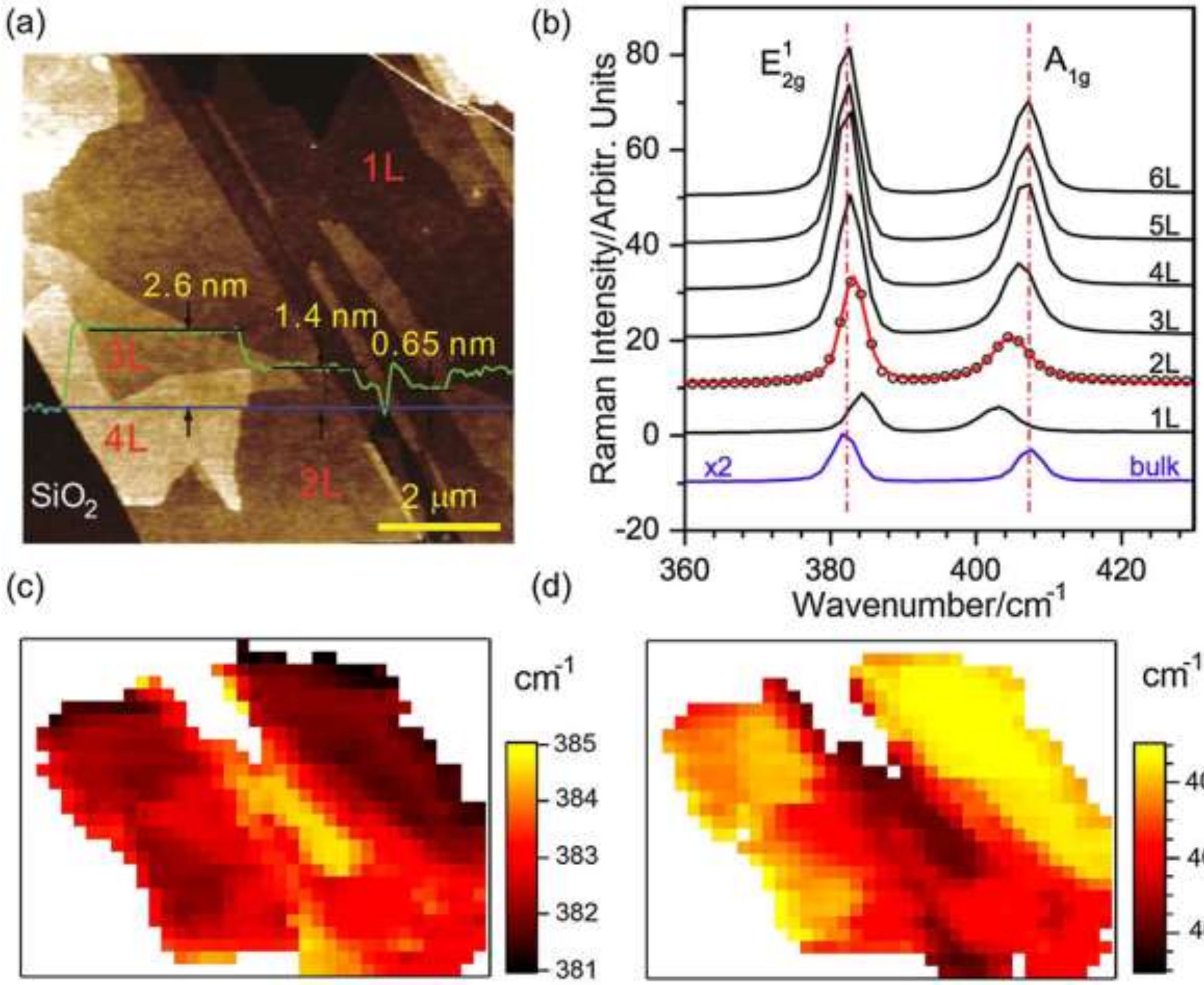

Fig 7

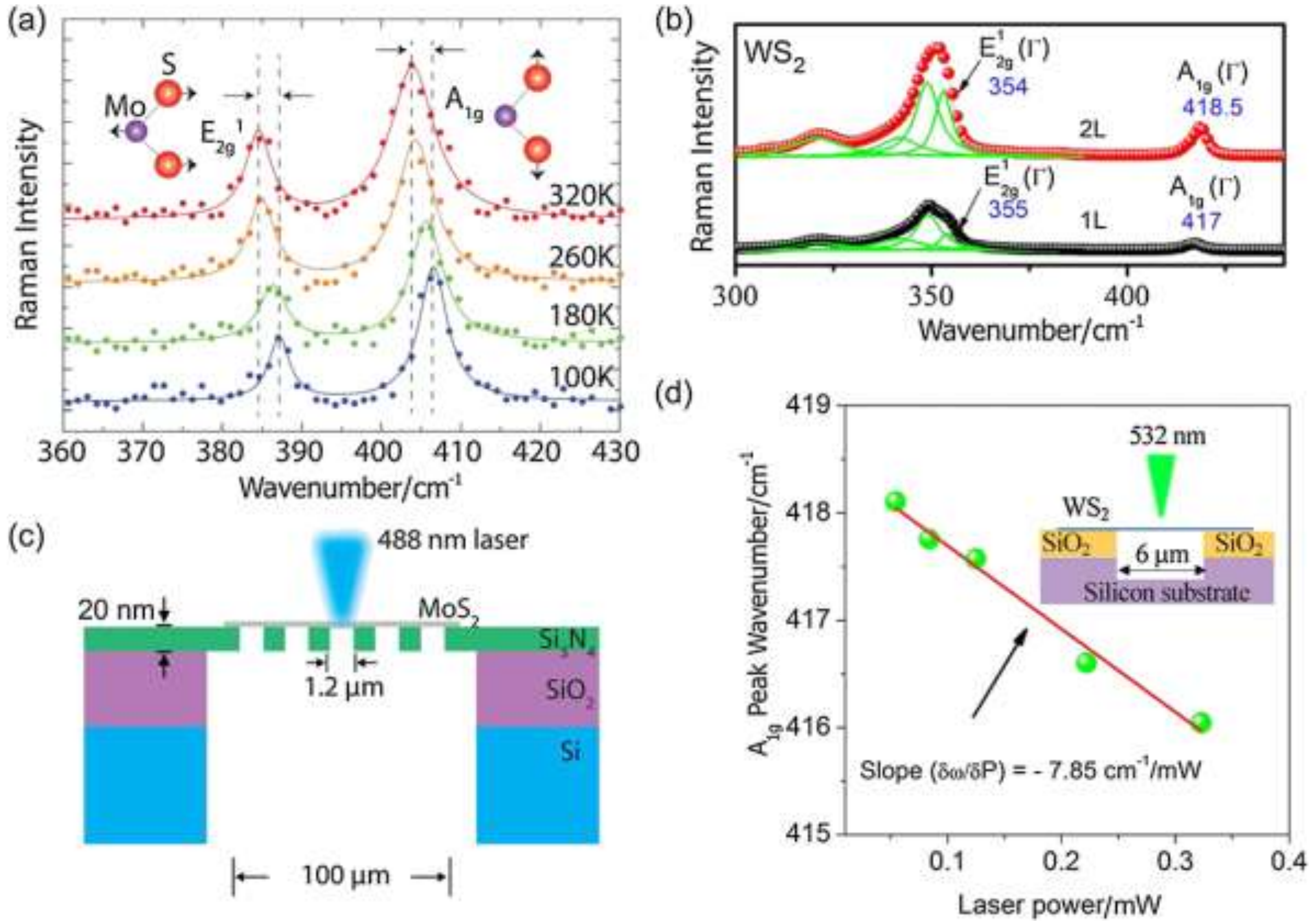
(a)
Raman Intensity
S
Mo
E2g 1
A1g
320K
260K
180K
100K
360 370 380 390 400 410 420 430
Wavenumber/cm-1
(b)
WS2
E1 2g (Γ)
354
A1g (Γ)
418.5
2L
E1 2g (Γ)
355
1L
A1g (Γ)
417
300 350 400
Wavenumber/cm-1
(c)
488 nm laser
20 nm
MoS2
Si3N4
SiO2
Si
1.2 μm
100 μm
(d)
A1g Peak Wavenumber/cm-1
419 418 417 416 415
532 nm
WS2
SiO2
6 μm
SiO2
Silicon substrate
Slope (δω/δP) = - 7.85 cm-1/mW
0.1 0.2 0.3 0.4
Laser power/mW

Fig 8